\documentclass[aps,prl,twocolumn,floatfix,superscriptaddress]{revtex4}
\usepackage{amsmath}
\usepackage{amssymb}
\usepackage{graphicx}
\usepackage{dcolumn}
\usepackage{bm}

\begin{document}

\title{Half-Integer Filling Factor States in Quantum Dots}

\author{A. Harju}
\email[Electronic address:\;]{ari.harju@hut.fi}
\affiliation{Laboratory of Physics, Helsinki University of Technology,
P.O. Box 1100, FI-02015 HUT, Finland}
\author{H. Saarikoski}
\affiliation{Kavli Institute of NanoScience, Delft University
of Technology, 2628 CJ Delft, the Netherlands}
\affiliation{Laboratory of Physics, Helsinki University of Technology,
P.O. Box 1100, FI-02015 HUT, Finland}
\author{E. R\"as\"anen}
\affiliation{Institut f{\"u}r Theoretische Physik,
Freie Universit{\"a}t Berlin, Arnimallee 14, D-14195 Berlin, Germany}
\affiliation{Laboratory of Physics, Helsinki University of Technology,
P.O. Box 1100, FI-02015 HUT, Finland}

\date{\today} 

\begin{abstract}
Emergence of half-integer filling factor states, such as $\nu$=5/2 and
7/2, is found in quantum dots by using numerical many-electron
methods. These states have interesting similarities and differences
with their counterstates found in the two-dimensional electron gas.
The $\nu=1/2$ states in quantum dots are shown to have high overlaps
with the composite fermion states. The lower overlap of the Pfaffian
state indicates that electrons might not be paired in quantum dot
geometry.  The predicted $\nu=5/2$ state has high spin polarization
which may have impact on the spin transport through quantum dot
devices.  PACS: 71.10.-w, 73.21.La, 73.43.-f, 85.35.Be
\end{abstract}

\maketitle

The fractional quantum Hall state at filling factor $\nu=5/2$ in the
two-dimensional electron gas (2DEG) is widely thought to be a
Moore-Read state of p-wave paired electrons with unusual topological
properties~\cite{read}. The state has recently attracted much
theoretical interest together with other half-integer filling factor
states, mainly due to proposals for using it in quantum
computing~\cite{freedman}. The 2DEG $\nu=5/2$ state is expected to
consist of two Landau levels (LL), where the filled lowest LL (0LL) is
spin-compensated and the half-filled next LL (1LL) is spin polarized
state described by a Pfaffian (PF) wave function~\cite{rezayi}. This
wave function explicitly describes the electron pairing which is
analogous to that in the BCS theory of superconductivity. Due to the
half-filling of 1LL at $\nu=5/2$, the properties of the $\nu=5/2$
state is usually analyzed using the $\nu=1/2$ state with an interaction
Hamiltonian corrected for the screening effects of the electrons in
the filled 0LL.

Unlike half-integer filling factor states in the 2DEG, the
corresponding states in quantum dots (QD's) have got almost no
attention until now (see Ref.~\onlinecite{emperador} for an
exception).  In this Letter we analyze the electronic structure of
QD's in magnetic fields and find new states which can be regarded as
finite size counterparts of the half-integer filling factor states in
the 2DEG.  Since the emergence of fractional quantum Hall states is
always a result of electron-electron interactions, we use numerical
methods which can take into account the complex correlation effects in
the system.  The results indicate analogies but also important
differences between the states in the half-integer filling factor
range in the 2DEG and QD's.  The mean-field method predicts that due
to the presence of an external confining potential in QD's, the
$\nu=5/2$ state appears as a mixture of $\nu=2$ and $3$.  We analyze
also the internal structure of the many-body state at $\nu=1/2$ in
QD's and compare it to the PF wave function and the composite fermion
(CF) model~\cite{cf}.  Unlike the CF model, which yields high overlaps
with the exact state, the PF wave function is found to poorly describe
the pure $\nu=1/2$ state and also the excited 1LL of the $\nu=5/2$
state in QD's. Therefore, we find no evidence of electron pairing near
half-integer LL fillings in QD's.  We finally discuss the impact of
the $\nu=5/2$ state on the ground state energetics and on the spin
transport through the QD and find that signatures of this state can be
found, e.g., in magnetization measurements or electron tunneling
experiments.

We define our QD system by an effective-mass Hamiltonian
for $N$ electrons
\begin{equation} H=\sum^N_{i=1}\left[\frac{({\mathbf p}_i+e{\bf A}
      )^2}{2 m^*}+V_c(r_i) +g^*\mu_B S_{z,i}\right] 
+ \sum_{i<j} \frac{e^2}{4\pi \epsilon r_{ij}} \ , 
\nonumber
\end{equation}
with material parameters for GaAs, i.e, $m^*=0.067$,
$\epsilon/\epsilon_0=12.4$, and $g^*-0.44$.  Unless stated otherwise,
we apply a parabolic confining potential, $V_c(r)=m^*\omega_0^2 r^2/2$
with a strength $\hbar \omega_0=5.7 N^{-1/7}\;{\rm meV}$.  The $N$
dependence in $\omega_0$ is introduced to keep the electron density in the dot
approximately constant. We solve the associated Schr\"odinger equation
using the mean-field spin-density functional theory (SDFT) and the
exact diagonalization (ED) method~\cite{clusters}.  We use the SDFT to
calculate the electronic structure of large quantum dots which are
beyond reach of exact many-body method. The internal structure of the
half-filled LL is then analyzed in detail with the ED method which is,
however, restricted to fairly low electron numbers.

We analyze the polarization and occupations of the LL's in large QD's
using the SDFT.  We focus here on the filling factor regime $\nu \geq
2$. One finds that the filling factor for the finite electron systems
is position dependent \cite{nu}. Even at integer filling factors
(where most physics can be understood on a single-particle level), the
2DEG and QD's have differences due to the overlapping of LL's in QD's
in low magnetic fields.  In QD's we identify the integer $\nu$ as
states, where the increasing magnetic field has emptied one of the
LL's completely of electrons. For example, $\nu=3$ corresponds to case
where magnetic field has emptied 2LL and the spin-compensated 1LL has
approximately one third of the electrons, the rest being at 0LL.  In
the 2DEG, 1LL is spin split at $\nu=3$ and has one spin channel fully
occupied.

\begin{figure}
\includegraphics[width=\columnwidth]{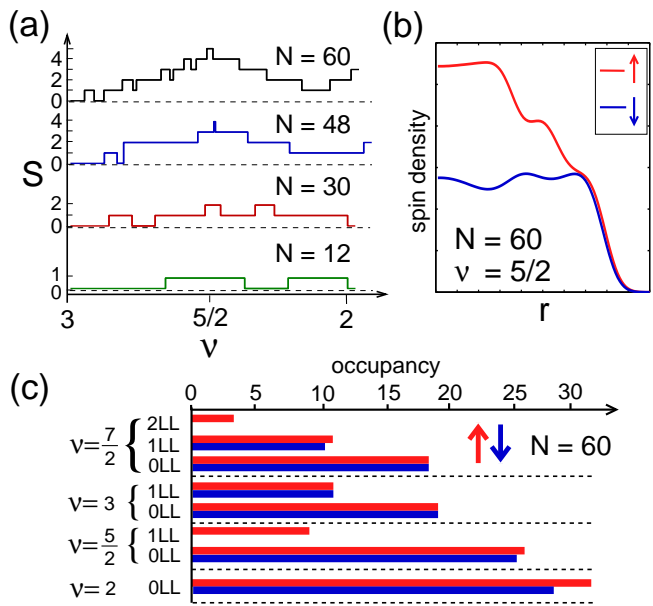}
\caption{(color online) (a) Ground-state spin of quantum dots with 
$N=12$, 30, 48, and 60 electrons, respectively.
The magnetic fields are scaled linearly to filling factor 
range $2\le \nu \le3$.
(b) Spin densities of the $N=60$ quantum dot at $\nu=5/2$.
The spin-polarized second Landau level gives rise to high spin 
polarization in the core region of the dot.
(c) Occupation of the single-electron 
(Kohn-Sham) states on different Landau levels in 
the $N=60$ quantum dot at $2\leq\nu\leq 7/2$.
}
\label{fig:1}
\end{figure}

Figure~\ref{fig:1}(a)
shows the calculated spin polarization $S=(N^\uparrow-N^\downarrow)/2$
of the ground state of QD's with $N=12$, 30, 48, and 60 electrons,
respectively.  Our starting point is the $\nu=3$ state discussed
above. When the magnetic field is increased electrons move from 1LL to
0LL.  This occurs mainly for spins anti-parallel to the magnetic
field.  We find a peak in $S$ between $\nu=2$ and $\nu=3$ where all
the spins anti-parallel to the magnetic field have fallen to 0LL
leaving 1LL totally spin polarized with angular momentum states $l =
-1,0,1,\dots,n_{\rm 1LL}-2$. Since the spin polarization of 1LL agrees
with the expected spin polarization of the Moore-Read state in the
2DEG, we identify this state as a finite-size counterpart of
$\nu=5/2$.  In addition, $\nu_{\rm avg}=2.54$ (see
Ref. \onlinecite{nu}) for this state when $N=60$.  The spin density of
$\nu=5/2$ state at $N=60$ shows that the spin polarization is
concentrated at the core of the dot [Fig.  \ref{fig:1}(b)]. This
result is in contrast with the structure of the 2DEG where the
single-particle states on 1LL are uniformly occupied, whereas in QD's
the occupation has a compact structure due to the external
confinement. Thus, the $\nu=5/2$ state in QD's can be interpreted to
be composed of two different filling factor domains, i.e., $\nu=2$ at
the edge and $\nu=3$ at the center. We find also an analogous state in
lower magnetic fields $\nu>3$ where the 2LL is spin polarized. We
identify this state as $\nu=7/2$. The Kohn-Sham occupations of LL's
for $\nu=2,\;5/2,\;3,$ and $7/2$, respectively, show the mechanism of
LL filling in QD's [Fig. \ref{fig:1}(c)]. Due to partial filling of
1LL, the $\nu=3$ state can also be interpreted as a combination of
$\nu=2$ and $\nu=4$ states.

Increasing the magnetic field beyond $\nu=5/2$ forces the
spin-polarized electrons in 1LL to fall one-by-one down to 0LL,
starting from the electrons with highest angular momenta. 1LL
maintains the spin polarization but the electrons in 0LL tend to
prefer spin compensated configuration and, as a consequence, there is
a partial relapse towards lower total spin polarization.  However,
near $\nu=2$ 0LL gets a small spin polarization in high electron
numbers, e.g., $S=1$ for $N=48$ and $S=2$ for $N=60$ [see
Fig.~\ref{fig:1}(a)].  This result is in accord with the results of
Ciorga and coworkers who found the collapse of the spin singlet state
in large QD's at $\nu=2$ (Ref.~\onlinecite{ciorga02}).

The results at $\nu=5/2$ can be generalized to larger particle numbers
by assuming a spin compensated 0LL and polarized 1LL.  The formula of
$\nu_{\rm 0LL}$ (see Ref. \onlinecite{nu}) gives $S \approx N/10$ for
the spin polarization at $\nu=5/2$, which is in a reasonable agreement
with the calculated maximum spin polarization of the two largest dots
in Fig. \ref{fig:1}(a).  As seen above in the case of $\nu=7/2$, we
find also evidence of analogous polarization of the highest occupied
LL in lower magnetic fields.  This suggests that the results can be
generalized to higher half-integer filling factor states ($\nu=9/2$
etc.) in sufficiently large QD's.

The predicted spin polarization of the states between $2\le \nu \le 3$
is interesting with a viewpoint on the spintronics in which electron
spin degree is exploited to create novel device concepts and new
functionality for electronic circuits.  It is assumed that the
alternation of states in the $2<\nu<3$ regime leads to the familiar
checkerboard pattern of conductance peak heights in the electron
tunneling through a QD device~\cite{keller,hitachi}.  However, the
details of the current patterns are not completely understood.  The
checkerboard behavior is usually explained assuming oscillations
between the spin singlet $S=0$ and triplet $S=1$ states~\cite{tarucha}
which can be qualitatively modeled within the constant-interaction
picture~\cite{stopa} with a phenomenological exchange term.  Our SDFT
calculations indicate that this model is sufficiently accurate in low
electron numbers [see, e.g., data for $N=12$ in Fig.  \ref{fig:1}(a)],
but there are significant deviations from it for $N\ge 20$.  The
constant-interaction model cannot take into account the correlation
effects which give rise to the high spin polarization near the
$\nu=5/2$ state in large QD's.  This might affect electron tunneling
probability through the dot for high electron numbers and, as a
consequence, leave a trace to the pattern of conductance peak heights.
However, the high probability transmission in the experiments occurs
mainly through the edge region which is not spin
polarized~\cite{keller}.  The transport through the center of the dot
gives rise to lower conductance peak heights, and evolution of the
ground state spin might leave a systematic pattern in the low
probability transmission peak heights.

We turn now into analysis of the exact many-body wave function of the
$\nu=1/2$ state.  We use the ED method on 0LL, and the
results are compared to two models, the PF and the CF wave
functions~\cite{cf}.  The PF wave function \cite{read} is defined as
\begin{equation}
 \Psi_{\rm PF} = \mathrm{Pf} \left( \frac{1}{z_i-z_j} \right) \prod_{i<j} (z_i-z_j)^2
\exp\left(-\frac 12 \sum_i r_i^2\right) \nonumber \ ,
\end{equation}
with angular momentum $L=N(N-1)-N/2$.  The value of the filling
factor at $\nu<1$, i.e., beyond the maximum-density-droplet (MDD)
domain is widely approximated to be $\nu_{\rm J}=L_{\rm MDD}/L$. This
definition is based on Jastrow states but the PF wave function has
$\nu_{\rm J}$ which is slightly higher than $1/2$. As the number of
particles in the system increases, $\nu_{\rm J}$ for the PF
approaches $1/2$ from above as $\nu_{\rm J}=1/2+1/2N+{O}(1/N)^2$.  The PF state is constructed for even
$N$, and we are computationally limited to $N \le 8$.  The CF wave
functions \cite{cf} used here are defined as
\begin{equation}
\Psi_{\rm CF} = \exp\left(-\frac 12 \sum_i r_i^2\right)
\Psi_0
\prod_{i<j} (z_i-z_j)^2 \nonumber 
\ ,
\end{equation}
where $\Psi_0$ is a determinant formed from the single-particle states
$\psi_{n,l}\propto z^{n+l} \partial^n$, with $l$ the angular
momentum and $n=0,1,2\dots$ is the LL index. The
derivative is with respect to $z$, and the resulting many-body state
has a total angular momentum $L=N(N-1)+\sum_{i=1}^N l_i$.

There are several possibilities for the single-particle occupations in
$\Psi_0$ for the states near $\nu_{\rm J}\approx 1/2$.  We consider
two different CF wave functions: CF1 that have the same angular
momentum as PF states, and CF2 that have $\nu_{\rm J}=1/2$, so that
$L=N(N-1)=2L_{\rm MDD}$.  The first type have single-particle quantum
numbers $(n,l)$ of the occupied states as $(0,i-1)$ and $(i,-i)$,
$i=1,\dots, N/2$, and in the second one the asymmetry in $l$ quantum
numbers is recovered by moving a particle from $(N/2,-N/2)$ to
$(1,0)$.  The CF2 wave function is constructed also for odd $N$.  The
numerical projection to 0LL (derivatives in $\Psi_0$) is doable up to
$N=6$.  From these wave functions, CF1 (and thus also PF) has an
angular momentum that corresponds to the ground-state of $N=4$, and
CF2 for $N=5$ and 6.  In Table~\ref{tab:o} we show the accuracy of the
three wave functions discussed above. PF is clearly less accurate than
CF which has very high overlaps.  The reasonably high overlap (around
.6) means, however, that PF cannot be an accurate excited state.
Therefore we find no evidence of electron pairing in the $\nu=1/2$
state.
\begin{table}
\caption{Overlaps of the trial wave functions with exact one for
different particle number $N$.}
\begin{ruledtabular}
\begin{tabular}{llll}
$N$ & PF & CF1 &  CF2\\ \hline
4   & 0.922124  & 0.999903 & 0.997978\\
5   & -                  & - &                 0.998921 \\
6   & 0.789996  & 0.993461 & 0.996096 \\
8   & 0.586370 & - & -
\end{tabular}
\label{tab:o}
\end{ruledtabular}
\end{table}

\begin{figure}
\includegraphics[width=.99\columnwidth]{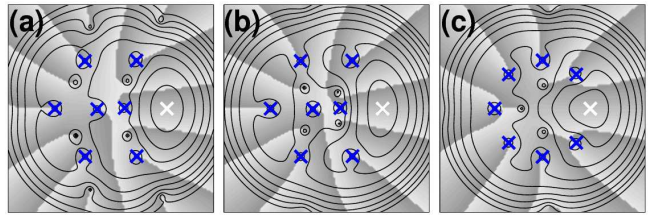}
\caption{(color online) Electron positions (crosses), conditional
electron densities (contours), and phases (gray-scale) of the $N=8$,
$L=52$ state from ED (a) and PF model (b and c).  We probe with the
rightmost electron and densities are on logarithmic scale.  The phase
changes from $\pi$ to $-\pi$ on the lines where the shadowing changes
from the darkest gray to the white.  In (a) and (c) electrons are on
the most probabe positions, and in (b) we use ED coordinates for PF.}
\label{fig:wf}
\end{figure}

Figure~\ref{fig:wf} shows the conditional wave functions \cite{clusters},
obtained by fixing $N-1$ electrons and moving the remaining electron,
for the PF and corresponding ED state. One can see that the most probable
positions are different in ED and PF and the ED conditional density
is more localized than the PF one. 

In the 2DEG the electrons in the half-filled 1LL can be described by a
$\nu=1/2$ state on 0LL having a screened electron-electron
interaction. This state is accurately modeled using a PF
wave function \cite{read,rezayi}.
The mean-field results above suggest that
because of the different structure of the $\nu=5/2$ states in QD's
and the 2DEG systems, the $\nu=5/2$ state
is not related to $\nu=1/2$ in parabolically confined QD's.
The 1LL in QD's is compact and
does not have the required angular momentum of the $\nu=1/2$ state.
One might assume that realistic inter-electron potentials,
which are modified by screening from the electrodes and
finite thickness of the sample, would induce electron pairing
as observed in numerical studies of the $\nu=5/2$ state in the 2DEG.
However, ED calculations with a screened electron-electron interaction
indicate that overlaps of the PF with the exact state increases
only marginally.
Therefore we find no evidence of electron pairing in the
analyzed half-integer filling factor states in QD's confined
by a parabolic external potential.

\begin{figure}
\includegraphics[width=\columnwidth]{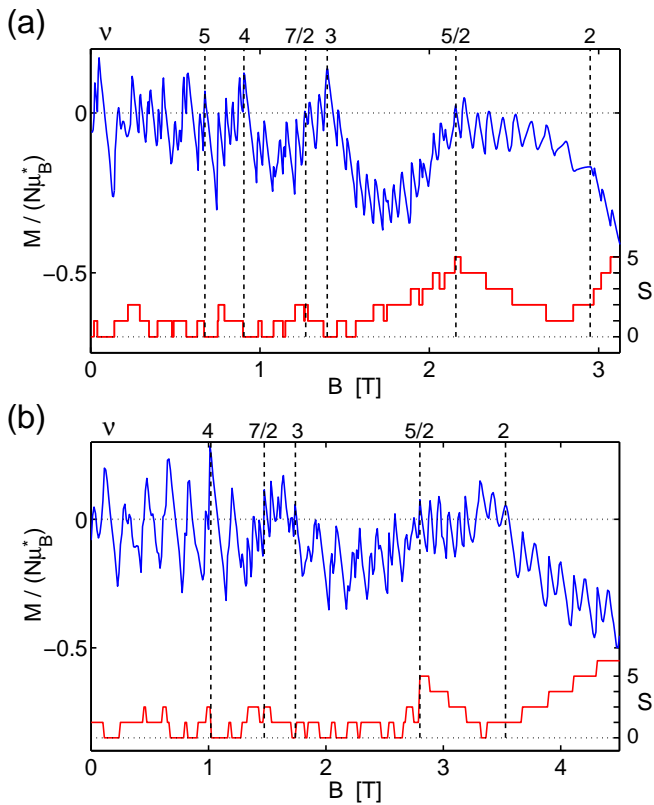}
\caption{(color online) Ground-state magnetization (upper curves) and
total spins (lower curves) of 60-electron quantum dots in parabolic
(a) and infinite-well (b) potentials.  The finite size counterparts of
the integer and fractional quantum Hall states are marked in the
figures (dashed lines).}
\label{fig:magnetization}
\end{figure}

The confinement potential in actual QD's may vary considerably from
the parabolic shape~\cite{bruce}. However, this fact seem not to
affect notably the appearance of the $\nu=5/2$ state in QD's.  Our
SDFT calculations for a 60-electron QD defined by an infinite
potential well of radius 150 nm yield the $\nu=5/2$ state similarly to
the parabolic case. As the main difference, the state is not separated
into two $\nu=2$ and $\nu=3$ domains in the infinite-well QD. Instead,
electrons on 0LL are strongly localized near the edge, whereas the
spin-up electrons on 1LL contribute in the mid-region leaving the
electron density in the core relatively flat for both spin types.  In
the ground-state energetics the behavior between different QD's is
similar, as we demonstrate in Fig.~\ref{fig:magnetization}
showing the SDFT result of the magnetization, $M=-\partial{E_{\rm
tot}}/\partial{B}$, for 60-electron parabolic and infinite-well QD's,
respectively.  The peaks in $M$ coincide with the integer filling
factors and constitute the finite size counterpart of the de Haas--van
Alphen effect.  In both cases, the $\nu=5/2$ state leaves cusps in the
magnetization, and the data shows also possible development of other
half-integer filling-factor states such as $\nu=7/2$ in lower magnetic
fields.  The results are consistent with the few direct magnetization
measurements of ensembles of large QD's which have shown a cusp-like
structure near $\nu=5/2$ (Ref.~\onlinecite{schwarz2}).  Visible
signatures are also expected to be found in the chemical potentials of
electron tunneling experiments.

To conclude, we predict emergence of states in quantum dots which can
be regarded as finite size counterparts of the half-integer filling
factor states in the 2D electron gas. The highest occupied Landau
level of these states is found to be spin-polarized and the formation
of them leaves signatures in the ground state energetics. The
numerical results indicate that these states share many of the
characteristics of their infinite size counterparts in the 2D electron
gas, although we find also major differences.  Due to observed
stability of these states we postulate that the formation of them is a
general property of finite fermion systems in external confinement.
In parabolic confining potential the composite fermion picture is
found to be a more accurate description of the $\nu=1/2$ state than
the Pfaffian wave function, which suggests that electrons are not
paired in this geometry.  The electron pairing, which is predicted to
occur in the two-dimensional electron gas, might be recovered in
quantum dots in the limit of very large electron numbers and with a
weak non-parabolic external confinement potential.

\acknowledgments

We thank S. Siljam\"aki for the CF wave function code, and J. Suorsa,
G.~E.~W. Bauer, M. Schwarz, S. Reimann and D. Grundler for fruitful
discussions.  This work was supported by the Academy of Finland
through the Centre of Excellence Program (2000-2011) and by the
NANOQUANTA NOE and the Finnish Academy of Science and Letters, Vilho,
Yrj{\"o} and Kalle V{\"a}is{\"a}l{\"a} Foundation (to E. R.).

\end{document}